\documentclass{article}
\usepackage{spconf, amsmath, graphicx}
\usepackage{subcaption, comment, multirow, multicol, makecell, amssymb, bm, textcomp, kotex, arydshln}
\usepackage{adjustbox}
\usepackage{mathtools}
\usepackage{cite, url}
\usepackage{tabularx, makecell}
\usepackage{algorithm}
\usepackage{algpseudocode}

\title{HM-Conformer: A Conformer-based audio deepfake detection system with hierarchical pooling and multi-level classification token aggregation methods}
\name{Hyun-seo Shin$^*$\thanks{$^*$Equal contribution}, Jungwoo Heo$^*$, Ju-ho Kim, Chan-yeong Lim, Wonbin Kim, and Ha-Jin Yu\sthanks{$^\dag$Corresponding author}\thanks{This work was supported by Institute of Information \& communications Technology Planning \& Evaluation(IITP) grant funded by the Korea government(MSIT) (No.RS-2023-00263037, Robust deepfake audio detection development against adversarial attacks)}}
\address{School of Computer Science, University of Seoul}
\begin{document}
\ninept
\maketitle
\begin{abstract}
Audio deepfake detection (ADD) is the task of detecting spoofing attacks generated by text-to-speech or voice conversion systems. 
Spoofing evidence, which helps to distinguish between spoofed and bona-fide utterances, might exist either locally or globally in the input features. 
To capture these, the Conformer, which consists of Transformers and CNN, possesses a suitable structure.
However, since the Conformer was designed for sequence-to-sequence tasks, its direct application to ADD tasks may be sub-optimal.
To tackle this limitation, we propose HM-Conformer by adopting two components: 
(1) Hierarchical pooling method progressively reducing the sequence length to eliminate duplicated information
(2) Multi-level classification token aggregation method utilizing classification tokens to gather information from different blocks.
Owing to these components, HM-Conformer can efficiently detect spoofing evidence by processing various sequence lengths and aggregating them.
In experimental results on the ASVspoof 2021 Deepfake dataset, HM-Conformer achieved a 15.71\% EER, showing competitive performance compared to recent systems.
\end{abstract}
\section{Introduction}
\label{sec:intro}
Recently, speech generation technologies such as voice conversion (VC) and text-to-speech synthesis (TTS) have become so sophisticated that their outputs cannot be distinguished from those of humans, and they are evolving dramatically\cite{min2021meta, casanova2022yourtts}.
Due to the risk of abusing these technologies in deepfake crimes or spoofing attacks, audio deepfake detection (ADD) has recently become a notable research area. 
In this flow, continuous efforts are being made to develop various countermeasure (CM) systems against spoofing. 
Among these efforts, CM systems based on deep neural networks (DNNs) have proven to be particularly effective, demonstrating outstanding performance\cite{wang2021comparative, jung2022aasist, liu2023leveraging}.

Over the last decade, CM systems that extract local features at the frame level and subsequently derive global features by aggregating them to the utterance level have revealed remarkable performance in ADD task. 
A prime example of this is LCNN-LSTM\cite{wang2021comparative}, which processes input features with light convolution neural network (CNN) and then reprocesses the extracted information with long short-term memory layers; this model has established itself as a benchmark in spoofing challenges\cite{yamagishi2021asvspoof}. 
More recently, models such as AASIST have emerged, leveraging a graph neural network as a back-end module to interpret the local features extracted by CNN, exhibiting superior results\cite{jung2022aasist}.
Furthermore, SE-Rawformer offers innovative approaches to CM systems by processing the CNN output with Transformers that operate on entire sequence lengths of the temporal axis\cite{liu2023leveraging}.

Liu et al.\cite{liu2023leveraging} argue that evidence of voice spoofing may exist at both the local and global levels, with examples including unnatural emphasis and intonation (at the local level) and excessive smoothing (at the global level). 
This assumption aligns well with the goal of CM systems that extract local features by employing convolutional layers and refine these features at the global level. 
From this perspective, we consider that the Conformer architecture\cite{gulati2020conformer}, which combines the Transformer specialized in extracting global features and the CNN specialized in exploring local features, would be well-suited for capturing evidence of voice spoofing.
Moreover, differing from traditional methods that explore local features first and then global features, the Conformer simultaneously explores local-global features by fusing the Transformer encoder with the convolution module. 
The architecture of Conformer can provide a standard for the importance of local information by considering the entire information.

%=====================
%       FIGURE
%=====================
\begin{figure*}[t]
\begin{center}
    \centering
    \includegraphics[width=\linewidth]{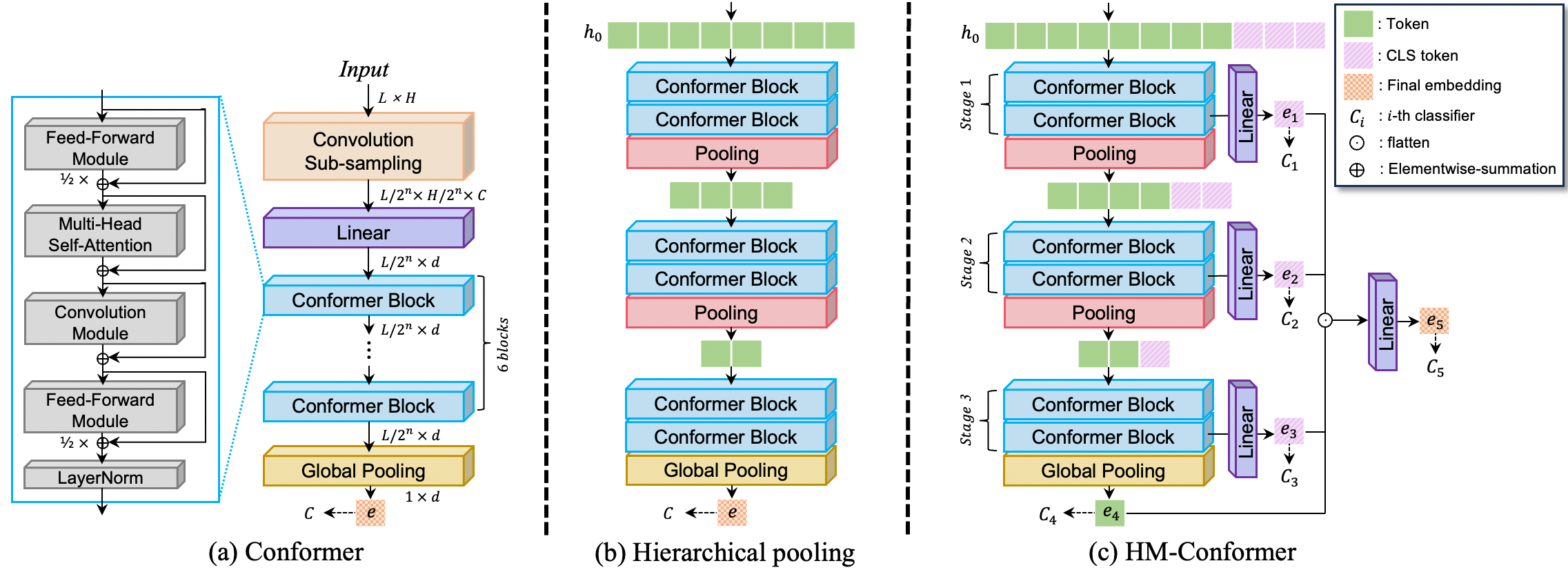}
    \vspace{-0.6cm}
    \caption{
    Overview of the Conformer-based architectures. 
    (a) depicts the overall architecture of the Conformer structure and the details of its blocks. 
    (b) and (c) illustrate the architecture with the hierarchical pooling method and the proposed HM-conformer with hierarchical pooling and MCA methods.
    }
\label{figure:conformer}
\end{center}
\vspace{-2.5em}
\end{figure*}

In this paper, we propose HM-Conformer by modifying the Conformer structure through hierarchical pooling and multi-level classification token aggregation (MCA) methods. 
The vanilla Conformer framework tends to carry duplicated information between frame-level features\cite{dai2020funnel}, since it was designed for automatic speech recognition that requires frame-level output (i.\,e., many-to-many task). 
On the other hand, in many-to-one scenarios such as classification, we hypothesize that conveying compact features is more advantageous than delivering overlapping features.
To reduce duplicated information, HM-Conformer applies a hierarchical pooling method which adds downsampling layers between Conformer blocks.
Furthermore, the strategy of utilizing information from various layers is widely recognized for enhancing the performance of classification tasks, including ADD task\cite{chen2021ur, zhang2020aret, wu2022attentional}. 
To this end, we devised MCA method to aggregate task-related information from various encoder blocks.
MCA method employs CLS tokens\cite{kenton2019bert}, which are effective in extracting information from the sequence of tokens, at each set of blocks to extract information from multiple encoders. 
Subsequently, the processed CLS tokens are individually trained through different classifiers and loss functions to extract task-related features.

HM-Conformer trained using the ASVspoof 2019 logical access training dataset achieved remarkable performance on the ASVspoof 2021 Deepfake detection task\cite{yamagishi2021asvspoof}. 
It achieved an equal error rate (EER) of 15.71\%, outperforming recent frameworks that did not employ ensemble techniques.

% 
% ----------------------------------------------------------
% Related Work - Conformer
% ----------------------------------------------------------
\section{Conformer}
\label{sec:Conformer}
Conformer\cite{gulati2020conformer} is an architecture proposed in the automatic speech recognition domain and has achieved superior performance compared to Transformer and CNN-based models\cite{koizumi2021df, chen2021continuous}. 
We attribute this performance to the fact that the Conformer adopts the advantages of both Transformer and CNNs by having a structure with a convolution module inserted within the Transformer encoder.
The Transformer is effective in modeling long-range global features by self-attention mechanism, while CNN is specialized for processing local features.
By fusing the two structures, the Conformer is able to capture both global and local features, which is suitable for detecting spoofing evidence scattered at a various range of scales.

As shown in Fig. \ref{figure:conformer} (a), Conformer consists of a convolutional sub-sampling and linear layer to tokenize the input, and Conformer blocks to process the tokens.
The convolution sub-sampling consists of $n$ 2D-convolution layers with a stride of 2 to adjust the sequence length of the input.
Following this, the vectors with flattened channel and frequency axes are processed into tokens $h_0\in\mathbb{R}^ {\frac{T}{2^n} \times d }$ through the linear layer.
The Conformer block has a structure with two feed-forward modules that wrap around a multi-head self-attention (MHSA) and a convolution module in the center.
If the output of the $i$-th Conformer block is defined as $h_i$, then the Conformer block can be represented by the following Equations:
\begin{gather}
    \widetilde{h}_{i-1}=h_{i-1} + \frac{1}{2}FFN\left(h_{i-1}\right), \\
    h^{\prime}_{i-1} = \widetilde{h}_{i-1} + MHSA\left(\widetilde{h}_{i-1}\right), \\
    h^{\prime\prime}_{i-1} = h^{\prime}_{i-1} + Conv\left(h^{\prime}_{i-1}\right), \\
    h_{i} = LayerNorm\left(h^{\prime\prime}_{i-1} + \frac{1}{2}FFN\left(h^{\prime\prime}_{i-1}\right)\right), 
\end{gather}
where $FFN$, $MHSA$, and $Conv$ refer to the feed-forward module, the multi-head self-attention module, and the convolution module, respectively.
Due to MHSA and the convolution module, the Conformer block can process global and local features separately at each layer.
Note that, unlike vanilla Conformer, we introduce global pooling using the SeqPooling\cite{li2022role} method to generate the final embedding $e \in \mathbb{R}^{1 \times d}$ from the output $h_6 \in\mathbb{R}^ {\frac{T}{2^n} \times d }$ to perform the ADD task.
Finally, as shown in Equation (5), the final embedding $e$ is input into the classifier $C$ to output a single scalar value $Score$, and this structure is used as a baseline.
\begin{gather} 
    Score = C(e) = \sigma\left(eW_1^c + b^c\right)W_2^c, 
\end{gather}
where $W_1^c \in \mathbb{R}^{d \times d/2}$, $W_2^c \in \mathbb{R}^{d/2 \times 1}$, and $b^c$ denote trainable parameters, and $\sigma$ is $Swish$ \cite{ramachandran2017swish} activation function.

\section{Proposed System: HM-Conformer}
\label{sec:HM-conformer}
In this study, we propose HM-Conformer for ADD task, which integrates two proposed methods into the Conformer structure: (1) hierarchical pooling method and (2) MCA method. 
The following two subsections and Fig. \ref{figure:conformer} (b), (c) describe the two proposed methods.

\subsection{Hierarchical pooling method}
\label{ssec:hierarchical pooling}
To improve the CM system using the Conformer, we noted that ADD task is a binary classification task, whereas Conformer was designed for sequence-to-sequence (seq-to-seq) tasks. 
In order to reduce the gap between the two tasks, we paid attention to research results that indicate tokens within the Transformer-based structure tend to become more similar to each other as they progress through encoders\cite{dai2020funnel}. 
Based on this observation, it has been argued in the image processing field that conveying compact information is more advantageous than providing duplicated information in many-to-one tasks such as classification\cite{pan2021scalable}. 
Inspired by this argument, we propose to apply a hierarchical pooling method that gradually downsamples the output of Conformer blocks to extract compact information for ADD task.
By decreasing the sequence length of the tokens, the Conformer block can propagate more condensed features to the subsequent blocks. 
In addition, the hierarchical pooling method offers one more advantage: it can reduce computational costs. 

Fig. \ref{figure:conformer} (b) illustrates the process of the hierarchical pooling method. 
First, the tokens $h_0$, output from the convolution sub-sampling and linear layers, are passed to the Conformer block. 
Then, the outputs of the Conformer block $h_i$ ($i \in \{2,4\}$) are processed through the pooling layer according to Equation (6). 
\begin{gather}
    \hat{h}_i = pooling\left(h_i; \gamma\right), 
\end{gather}
where $\gamma$ means the downsampling rate, which is fixed at 2 in this paper, and if $h_i \in \mathbb{R}^{T^{\prime} \times d}$, then $\hat{h}_i \in \mathbb{R}^{ \frac{T^\prime}{\gamma} \times d }$. 

% ----------------------------------------------------------
\begin{figure}[t]
\vspace{-1.3em}
\begin{algorithm}[H]
\caption{The algorithm of the HM-Conformer}\label{alg:cap}
\begin{algorithmic}[1]
\State $\gamma \gets 2$         \Comment{Downsampling rate}
\State $i \gets 1$
\State $j \gets 3$                      \Comment{Number of CLS tokens}
\State $E \gets [\cdot]$                \Comment{Embedding list}
\State $h_0 \gets \{ cls_1, cls_2, cls_3, t_1, ... , t_{T/2^n}\}$   \Comment{Add CLS tokens}
\While{$i \leq 6$}
\State $h_{i} \gets ConformerBlock(h_{i-1})$
\If{$i \in \{2, 4\}$ }                  \Comment{After 2, 4-th Conformer block}
    \State $CLS \gets h_{i}[ :j] $        \Comment{Separate CLS tokens}
    \State $E \gets E \cup \{CLS[0]\cdot W_j + b_j\} $       
    \State $\hat{h}_{i} = pooling\left(h_{i}[ j : ]; \gamma\right)$  \Comment{Hierarchical pooling}
    \State $h_{i} \gets \{ CLS[ 1: ] \} \cup \hat{h}_{i}$
    \State $j \gets j - 1$

\ElsIf{$i \in \{6\}$ }                  \Comment{After 6-th Conformer block}
    \State $CLS \gets h_{i}[ :j] $
    \State $g \gets SeqPool(h_{i}[ 1 : ])$  \Comment{Global-level token}
    \State $E \gets E \cup \{CLS[0]\cdot W_j + b_j, g\}$       
\EndIf
\State $i \gets i + 1$
\EndWhile
\State $e_5 \gets flatten(E)\cdot W_4 + b_4 $                   \Comment{Final embedding}

\end{algorithmic}
\end{algorithm}
\vspace{-2.5em}
\end{figure}
% ----------------------------------------------------------

\subsection{Multi-level classification token aggregation method}
\label{ssec:stage-cls token aggregation}
The approach of aggregating and processing outputs from various layers, known as multi-level feature aggregation, is known to enhance the performance of classification tasks\cite{chen2021mfanet, chen2021ur, desplanques2020ecapa}. 
From this perspective, utilizing features extracted from multiple Conformer blocks may be beneficial for ADD task.
However, lower layers of Transformer-based models are observed to process less task-relevant information\cite{yu2022auxiliary}, making the direct use of outputs from Conformer blocks potentially inefficient. 
Taking these characteristics into account, we propose MCA method that extracts task-relevant features from multiple blocks. 

MCA is a method that extracts task-related information by training the CLS token, a widely used feature extraction technique in Transformer-based models for classification tasks, through auxiliary losses.
CLS tokens are learnable vectors inserted at the beginning of a token sequence that serve as an aggregated representation of the sequence through MHSA modules \cite{kenton2019bert}. 
Then, the aggregated representation can be utilized for classification tasks. 
Given this consideration, MCA method adds CLS tokens to the input sequence, and each set of Conformer blocks (called stage) has its own classifier and a loss function. 
Therefore, the lower block can be trained with more strong task-relevant supervision.
Furthermore, MCA method can aggregate more discriminative features by adapting CLS tokens for each stage; when applied with the hierarchical pooling method, it becomes feasible to gather even more discriminative features from token sequences with diverse time scales. 

Fig. \ref{figure:conformer} (c) depicts HM-Conformer where we applied the hierarchical pooling and MCA methods. 
Algorithm \ref{alg:cap} shows the process of HM-Conformer.
First, three CLS tokens which are randomly initialized vectors (pink dotted boxes in Fig. \ref{figure:conformer} (c)) are added to the input token sequence $h_0$ as presented in Equation (7). 
\begin{gather}
    {h}_0 = \{ cls_1, cls_2, cls_3, t_1, ... , t_{T/2^n}\},
\end{gather}
where $cls_i\in\mathbb{R}^d$ and $t_i\in\mathbb{R}^d$ denote CLS tokens for each stage and the tokens from input, respectively.
These tokens are processed by the Conformer blocks, enabling CLS tokens to aggregate information from other tokens. 
The refined CLS tokens are separated before entering the pooling layer and then used to produce the final embedding with a global-level token (the lowest green box). 
Each of the CLS tokens, global-level token, and final embedding are transformed into an embedding $e_k$ through a classifier, as shown in Algorithm \ref{alg:cap}.
Subsequently, each embedding processed through its own OC-Softmax\cite{zhang2021one} loss function $L_{os}$ to calculate the final loss $L$: 
\begin{gather}
    {Score}_k = C_k(e_k) = \sigma\left(e_kW_1^k + b^k\right)W_2^k, \\
    L_{os_k} = \frac{1}{N}\sum_{j=1}^{N}\log\left(1+\exp^{\alpha\left(m_{y_j}-\hat{w}_k{Score}_k\right)\left(-1\right)^{y_j}} \right), \\
    L = \sum_{k=1}^5 w_k L_{oc_k}\left({Score}_k\right),    
\end{gather}

where $\hat{w}_k$, $\alpha$, and $m_{y_j}$ denote a trainable single parameter, a scale factor, and margins, respectively.
$y_j$ means labels that $y_j = 0$ for bona-fide and $y_j = 1$ for spoofed, and the hyper-parameter $w_k$ denotes weight of $L_{oc_k}$. 
Note that during inference, only the $Score_5$ processed from the final embedding $e_5$ is used.

%================
%     TABLE
%================
\begin{table}[t]
\caption{
 Comparison of EER (\%) performance of the ASVspoof 2021 DF task evaluation in various frameworks. (*: our implementation)
}
\vspace{-0.5em}
\centering
\resizebox{0.8\linewidth}{!}{
\label{table:table1}

\begin{tabular}{lc|c}
\Xhline{2\arrayrulewidth}
\multicolumn{1}{c}{\multirow{2}{*}{\textbf{Frameworks}}} && \multirow{2}{*}{\textbf{EER (\%)}} \\&& \\
\Xhline{2\arrayrulewidth}
\rule{0in}{3ex}
LFCC-LCNN (Wang et al., 2021\cite{wang2021comparative})    && 23.48   \\ 
\,SE-Rawformer* (Liu et al., 2023\cite{liu2023leveraging})     && 21.65   \\ 
\,AASIST (Jung et al., 2022\cite{jung2022aasist})           && 20.04   \\ 
\,SFR-CNN (Yamagishi et al., 2021\cite{yamagishi2021asvspoof})           && \,19.22   \\ [1ex]
\hline
\rule{0in}{3ex}
Conformer (baseline)   && 18.91   \\ 
\,\textbf{HM-Conformer}                   && \,\textbf{15.71}    \\ [1ex]
\Xhline{2\arrayrulewidth}
\end{tabular}}
\vspace{-1em}
\end{table}

% ----------------------------------------------------------
% Experimental setup
% ----------------------------------------------------------
\section{Experimental setup}
\label{sec:experimental setup}

\subsection{Datasets and evaluation metric}
\label{ssec:datasets}
We used the training and development partitions of the ASVspoof 2019 logical access task datasets for model training. 
This training dataset consists of 5,128 bona-fide and 45,096 spoof utterances generated by six spoofing attack algorithms. 
Model evaluation was performed on the evaluation partitions of the ASVspoof 2021 deepfake (DF) task dataset. 
The DF evaluation dataset comprises 611,829 samples of 100 spoofed and bona-fide utterances combinations. 
We compared the performance of the models based on the Equal Error Rate (EER), the official metric for the ASVspoof 2021 challenge.

\subsection{Implementation details}
\label{ssec:implementation details}
We employed the OC-Softmax\cite{zhang2021one} loss function with hyper-parameters $\alpha=20$, $m_0=0.9$, and $m_1=0.2$. 
To augment training data, we employed media codec\footnote{media codec: aac, flac, mp3, m4a, wma, ogg, wav} augmentation, speed perturbation set to 0.9 and 1.1, SpecAugment\cite{park2019specaugment} for randomly masking 0 to 20 frequency axes, and colored additive noise augmentation with signal-to-noise ratios randomly selected between 10 and 40. 
For input features, we used 400 frames of the 120-dimensional linear frequency cepstral coefficients (LFCCs), encompassing a window length of 20ms, a hop size of 10ms, a 512-point FFT, a linearly spaced triangle filter bank of 40 channels, and delta and delta-delta coefficients. 
We utilized a batch size of 240 and the Adam\cite{kingma2014adam} optimizer with $\beta1 = 0.9$ and $\beta2 = 0.999$. 
Readers can access our codes at GitHub\footnote{https://github.com/talkingnow/HM-Conformer}.

%=====================
%       FIGURE
%=====================
\begin{figure}[t]
\begin{center}
    \centering
    \includegraphics[width=0.85\linewidth]{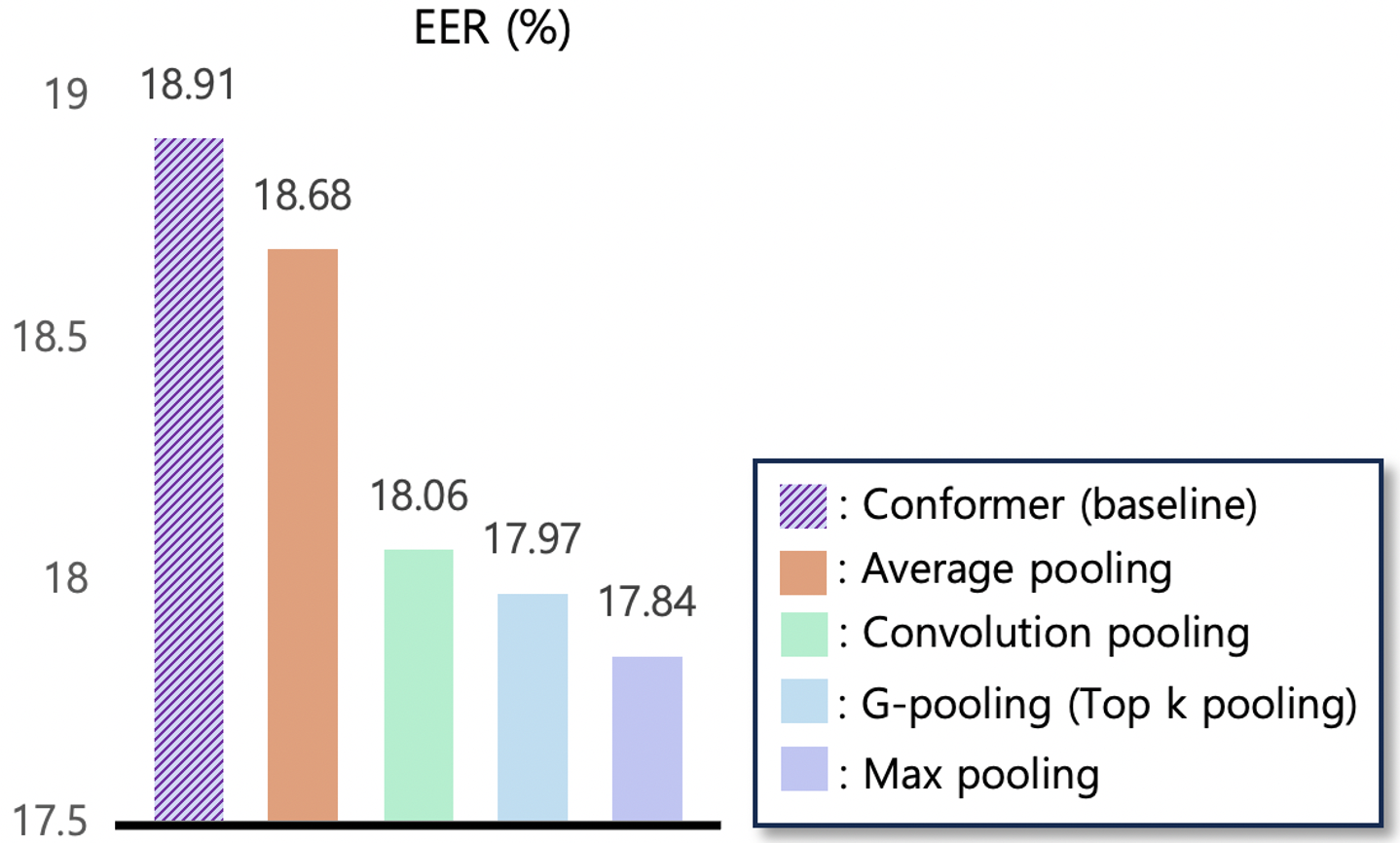}
    \vspace{-0.2cm}
    \caption{Comparison of EER (\%) performance between various pooling layers applied to hierarchical pooling method. Convolution pooling was performed using a convolution layer with a kernel size seven and a stride four. G-pooling adopted the method of Gao et al,\cite{gao2019graph}.}
\label{figure:pooling_compare}
\end{center}
\vspace{-3em}
\end{figure}

% ----------------------------------------------------------
% Result
% ----------------------------------------------------------
\section{Results}
\label{sec:results}
\subsection{Comparison with recent CM systems}
Table \ref{table:table1} shows the comparison of the performance of the HM-Conformer with recently proposed single CM frameworks on the ASVspoof 2021 DF evaluation. 
The baseline framework (Conformer) reveals an EER of 18.91\%, which is improved performance than other frameworks, thereby validating its potential in ADD. 
The proposed HM-Conformer achieved an EER of 15.71\%, representing an approximately 16\% improvement over the baseline. 
These results indicate that we successfully adapted the vanilla Conformer framework for the ADD task using our proposed hierarchical pooling and MCA methods.
%These results imply that conventional Conformer has potential, but we successfully migrated many-to-many models to many-to-one scenarios with hierarchical pooling and stage-CLS token aggregation methods. 

\subsection{Validation of the hierarchical pooling method}
To validate the effectiveness of the hierarchical pooling method, we confirmed the performance variation when applying different types of pooling layers, as displayed in Fig. \ref{figure:pooling_compare}. 
In our experiments, all employed pooling strategies yielded superior performance compared to the baseline. 
These results demonstrate that conveying condensed features is reasonable for addressing the ADD task. 
Meanwhile, there is one further notable observation in Fig. \ref{figure:pooling_compare}. 
In previous studies on ADD task, pooling mechanisms that select more significant representations from extracted features, such as max-pooling, are often employed in building CM systems and show outstanding performance \cite{jung2022aasist, wang2021comparative}.
Consistent with prior works, max and top-k pooling derived better performance than other pooling techniques in our experiments.  \\
%Our interpretation of these results is that considering the characteristic of the task, which requires capturing spoofing evidence, the procedure of selecting features is well-suited for this task. 

%================
%     TABLE
%================
\begin{table}[t]  
\caption{
Comparison of EER performance when changing the weights $w_k$ of the loss function.
}
\vspace{-0.5em}
\centering
\resizebox{0.95\linewidth}{!}{
\label{table:w_compare}
\begin{tabular}{l|ccccccccccc|c}
\Xhline{2\arrayrulewidth}
\multirow{2}{*}{\textbf{No.}} && \multirow{2}{*}{\textbf{$w_1$}} & & \multirow{2}{*}{\textbf{$w_2$}} & &\multirow{2}{*}{\textbf{$w_3$}} & &\multirow{2}{*}{\textbf{$w_4$}} & &\multirow{2}{*}{\textbf{$w_5$}} && \multirow{2}{*}{\textbf{EER (\%)}} \\ &&&&&&&&&&&& \\
\Xhline{2\arrayrulewidth}
\,\#1 &&&  & & & &&&&& &  \,17.84  \\ 
\hline
\rule{0in}{3ex}
\#2 && 1 &:& 1 &:& 1 &:& 1 &:& 6 &&  17.07  \\ 
\,\#3 && 1 &:& 1 &:& 2 &:& 3 &:& 4 &&  17.03  \\ 
\,\#4 && 1 &:& 1 &:& 1 &:& 1 &:& 1 && 16.06  \\ 
\,\textbf{\#5} && \textbf{4} &\textbf{:}& \textbf{3} &\textbf{:}& \textbf{2} &\textbf{:}& \textbf{1} &\textbf{:}& \textbf{1} &&  \textbf{15.71} \\ 
\,\#6 && 6 &:& 1 &:& 1 &:& 1 &:& 1 &&  \,15.72  \\ [1ex]
\Xhline{2\arrayrulewidth}
\end{tabular}}
\vspace{-0.5em}
\end{table}

%================
%     TABLE
%================
\begin{table}[t]
\caption{
Comparison of EER performance for ablation study using the sub-set of $e_k$.
$e_1, e_2$, and $ e_3$ denote CLS token from $k$-th stage, and $e_4$ means global-level token. 
(Ratio of loss weights 4:3:2:1)
}
\vspace{-0.5em}
\centering
\resizebox{0.9\linewidth}{!}{
\label{table:ablation}
\begin{tabular}{l|ccc|ccc|ccc|ccc|c}
\Xhline{2\arrayrulewidth}
\multirow{2}{*}{\textbf{No.}} & &\multirow{2}{*}{} \multirow{2}{*}{\textbf{$e_1$}}&&\multirow{2}{*}{}   &\multirow{2}{*}{} \multirow{2}{*}{\textbf{$e_2$}} &&\multirow{2}{*}{}  &\multirow{2}{*}{} \multirow{2}{*}{\textbf{$e_3$}}&&\multirow{2}{*}{}  &\multirow{2}{*}{} \multirow{2}{*}{\textbf{$e_4$}} &\multirow{2}{*}{} & \multirow{2}{*}{\textbf{EER (\%)}}  \\ &&&&&&&&&&&&& \\
\Xhline{2\arrayrulewidth}
\rule{0in}{3ex}
\#1 && √ &&&  &&&  &&& √ && 17.41    \\ 
\,\#2 &&  &&& √ &&&  &&& √ &&  16.59    \\ 
\,\#3 &&  &&&  &&& √ &&& √ &&  17.38    \\ 
\,\#4 && √ &&&  &&& √ &&& √ && 16.92   \\ 
\,\#5 &&  &&& √ &&& √ &&& √ && 17.74    \\ 
\,\#6 && √ &&& √ &&& √ &&&  && \,16.71    \\ [1ex]
\hline
\,\#7 && √ &&& √ &&& √ &&& √ && \,15.71   \\
\Xhline{2\arrayrulewidth}
\end{tabular}}
\vspace{-1.5em}
\end{table}

\subsection{Effectiveness of MCA method}
Table \ref{table:w_compare} shows the results of the Conformer with MCA method under various conditions of the $w$ ratio. 
Experiment \#1 shows the EER of the baseline, which is the Conformer framework with a max pooling as described in subsection 5.2.
Compared to the baseline, all experiments with MCA achieved superior performance as depicted in \#2$\sim$\#6. 
Based on these results, we concluded that our proposed MCA method could improve the Conformer's performance in the ADD task by transmitting appropriate information for detecting spoofing evidence. 
We also found that increasing the loss weight for the lower layers resulted in better performance than vice versa (\#2, \#3 vs \#5, \#6). 
In the end, by adjusting the MCA with $w$ ratio 4:3:2:1:1 to the baseline, we attained the best performance HM-Conformer, which shows an EER of 15.71\%.

\subsection{Ablation study of MCA method}
In Table \ref{table:ablation}, we performed a token ablation experiment on the best HM-Conformer to prove that all of the different stages of information are valid. 
We observed that the performance of experiments \#1$\sim$\#6, excluding some elements, decreased compared to experiment \#7, which used all elements.
The results of these experiments suggest that all CLS tokens and the global-level token carry information regarding the spoofing evidence to the final embedding and that diverse discriminative information is significant for performance improvement.

% ----------------------------------------------------------
% Conclusion
% ----------------------------------------------------------
\section{Conclusion}
\label{sec:conclusion}
In this study, we propose the HM-Conformer, a spoofing CM system, by modifying the Conformer structure through hierarchical pooling and multi-level classification token aggregation methods. 
The hierarchical pooling method, which can narrow the gap between seq-to-seq tasks and classification tasks, extracts compressed information suitable for the ADD task by reducing the sequence length of the Conformer block. 
MCA method enables the model to discern spoofing evidence from diverse sequence lengths at varying time compression levels.
We verified that these two methods can enhance the Conformer, resulting in competitive performance in the ADD task when compared to modern frameworks.

% References should be produced using the bibtex program from suitable
% BiBTeX files (here: strings, refs, manuals). The IEEEbib.bst bibliography
% style file from IEEE produces unsorted bibliography list.
% -------------------------------------------------------------------------
\bibliographystyle{IEEEbib}
\bibliography{strings,refs}

\end{document}